\documentclass[12pt,a4paper,twoside]{article}
\usepackage{epsfig}
\usepackage{graphicx}
\usepackage{amsmath}
\usepackage[widemargins]{a4}

\begin{document}
\date{September 30, 2002}
\title{The problem of hierarchies and family replications in an 
 extension of
quantum field theory }
\author{Francesco Caravaglios \\
%EndAName
Dipartimento di Fisica, Universit\`a di Milano, \textit{and } INFN sezione
di Milano}
\maketitle

\begin{abstract}
We will show that an extension of quantum field theory, recently proposed to
solve the hierarchy problem, can give an elegant explanation of quark/lepton
family replications. This scenario prefers fermion mass models based on \ a
family permutation discrete symmetry.
\end{abstract}

\section{Introduction: a theory of identical families}

In quantum mechanics, a particle is described by three real numbers $%
(x_{1},x_{2},x_{3})$ that correspond to the space position of the particle.
To describe a system with several identical particles, we need to introduce a
function $\psi (x)$ \ of the space coordinates above. This function is
called a field, and quantum field theory or second quantization is the
theory that quantizes it . Now suppose that we have a system with several
identical fields. This could be the system of three identical fields $\psi
_{e}(x),\psi _{\mu }(x),\psi _{\tau }(x)$, but with different masses: the
electron, the muon and the tau. Repeating the above argument, one can
imagine that a theory of several identical families needs the introduction of
a new mathematical object to represent such a physical system: a function of
the field $\psi (x)$, \textit{i.e.} a functional $S[\psi (x)]$. We have
recently shown \cite{primo}, that the quantization of \ functionals gives us
an extension of quantum field theory. Such a theory has been proposed \ to
solve the hierarchy problem. Here we will see how this idea can nicely fit
three generations of fermions and correctly realizes fermion masses and
mixings.

\section{A model with three families}

In this section we will present a model \ with three fermion families. To
simplify the notation, we will \ ignore the gauge structure, since it is
irrelevant for the discussion below. 

We have already discussed \ the case of
scalar fields $\phi $, in the paper \cite{primo}. Here we repeat a similar
argument to introduce fermions. We remind the anti-commutation rules for a
fermionic field operator $\hat{\psi}(x)$%
\begin{equation}
\{\hat{\psi}(x),\ \hat{\psi}^{\dagger }(y)\}=\delta ^{3}(x-y).
\end{equation}
We can give a functional representation to these operators. In fact, in
quantum field theory any quantum field state can be represented by a functional%
\footnote{%
This functional is the second quantization analogue of the particle wave
function in first quantization \cite{primo}.} $S[\psi (x)]$, where $\psi (x)$ is a
Grassman variable, function of the three space coordinates $x$. The action
of the operators $\hat{\psi}(x)$ and $\hat{\psi}^{\dagger }(x)$ onto
ordinary quantum field states, can be represented by the function $\psi (x)$
and the functional derivative $\delta \,/\delta \psi (x)$, acting on this
functional $S[\psi ]$  (the wave functional of the state) 
\begin{equation}
\begin{tabular}{l}
$\hat{\psi}(x)|S\rangle \ \Leftrightarrow\  \psi (x)\,\ S[\psi ]$ \\ 
$\hat{\psi}^{\dagger }(x)|S\rangle \ \Leftrightarrow \  $ $ \,\frac{\delta }{%
\delta \psi (x)}\,\ S[\psi ]$.%
\end{tabular}
\label{subst}
\end{equation}
In fact they satisfy the anti-commutation rules 
\begin{equation}
\{\psi (x),\ \frac{\delta }{\delta \psi (y)}\}=\delta ^{3}(x-y).
\label{anti}
\end{equation}
The Hamiltonian of a free Dirac field, after the substitution (\ref{subst}),
 becomes 
\begin{equation}
H=\int d^{3}x\,\psi ^{\dagger }(x)\ (-i\gamma ^{0}\mathbf{\gamma }\cdot 
\mathbf{\nabla +}m\,\gamma ^{0})\frac{\delta }{\delta \psi ^{\dagger }(x)}+%
\text{h.c.}  \label{hami2}
\end{equation}
This is the Hamiltonian in the functional representation. In the Schr\"{o}%
dinger picture,  a quantum field state is represented by 
a wave functional $S[\psi,t]$,  whose time evolution, in terms of the 
time variable $t$, can be
computed solving the Schr\"{o}dinger functional equation 
\begin{equation}
i\frac{\partial }{\partial t}S[\psi ,t]=H\,\ S[\psi ,t]  \label{eq:schro}
\end{equation}
with $H$ given by the (\ref{hami2}).
The time evolution equation (\ref{eq:schro}), is the quantum field theory
analogue of the Schr\"{o}dinger equation in first quantization. We stress that we 
have not yet 
introduced any new physical concept.

Now we briefly repeat the arguments discussed in \cite{primo}, 
leading to the   extension of quantum field theory.     
 As  explained in \cite{primo}, 
one can derive this Schr\"{o}dinger equation (\ref{eq:schro}), from
 the stationarity condition  of a new action $\mathcal{A}$, written in terms of
functionals 
\begin{equation}
\mathcal{A=}\mathop{\displaystyle \int}\mathcal{D}\psi \,S^{\dagger }[\psi
,t]\left( i\frac{\partial }{\partial t}+H\right) S[\psi ,t]\,\,dt+\text{h.c.}
\end{equation}

where $H$ is given in (\ref{hami2}) and the integral \ $\int \mathcal{D}\psi $
is performed in the functional sense. 
The classical action $\mathcal{A}$ can be quantized \cite{primo}. In
particular we have that $S$ and $S^{\dagger }$ are replaced with functional
field operators that satisfy the anti-commutation rules 
\begin{equation}
\{iS^{\dagger }[\psi ],S[\psi ^{\prime }]\} = \delta ^{\infty }[\psi
-\psi ^{\prime }]  \label{eq:anti1}
\end{equation}
where $\delta ^{\infty }[\psi -\psi ^{\prime }]$ is a functional Dirac delta 
\begin{equation}
\int \mathcal{D}\psi \,\delta ^{\infty }[\psi -\psi ^{\prime }]\,G[\psi
]=G[\psi ^{\prime }].
\end{equation}
As expected,  the Hamiltonian obtained from the
action $\mathcal{A}$, is written in terms of a functional integration
\begin{equation}
\mathcal{H}=\mathop{\displaystyle \int}\mathcal{D}\psi \,S^{\dagger }[\psi
]\,\ H\,\ S[\psi ]\,\,.  \label{ham}
\end{equation}
Using the (\ref{eq:anti1}), one can verify that the operator 
\begin{equation}
\mathcal{N=}\mathop{\displaystyle \int}\mathcal{D}\psi \,S^{\dagger }[\psi
]\,\ S[\psi ]
\end{equation}
satisfies the commutation rules 
\begin{eqnarray}
\left[ \mathcal{N},S[\psi ]\right]  &=&-S[\psi ] \\
\left[ \mathcal{N},S^{\dagger }[\psi ]\right]  &=&S^{\dagger }[\psi ]. 
\notag
\end{eqnarray}
\ $\mathcal{N}$ counts the number of fields. $S$ and $S^{\dagger }$ are
annihilation and creation functional field operators. So, we can exploit $%
S^{\dagger }$ to build different states of the Hilbert space, starting 
 from the vacuum state, that satisfies by definition 
\begin{equation}
S[\psi ]\,\ |0\rangle =0.  \label{vuot}
\end{equation}
Since our purpose is to build a model with three families,  we consider
 a state $|F_3\rangle$
with three creation operator $S^{\dagger }$, and study  the  evolution  
with respect the time
variable $t$:
\begin{equation}
|F_{3},t\rangle =\mathop{\displaystyle \int}\mathcal{D}\psi _{1}\mathcal{D}%
\psi _{2}\,\mathcal{D}\psi _{3}\,\ F_{3}[\psi _{1},\psi _{2},\psi _{3},t]\,\
\ S^{\dagger }[\psi _{1}]\,\ S^{\dagger }[\psi _{2}]\ S^{\dagger }[\psi
_{3}]|0\rangle .  \label{vuoto}
\end{equation}
This is a state with three fields $\psi _{1}(x),\psi _{2}(x),\psi _{3}(x)$,
and the functional $F_{3}[\psi _{1},\psi _{2},\psi _{3},t]$ unambiguously
defines the specific physical state of the  system. The normalization 
constant of the functional $F_3$ is chosen in order to guarantee 
$\langle F_3 |F_3\rangle=1$ (see also \cite{primo}). Since $\mathcal{N}$
commutes with the Hamiltonian (\ref{ham}), the number of fields is constant.
Exploiting  the Hamiltonian  (\ref{ham}), we can derive 
 the  time evolution of the functional $F_3$. Namely, we
have the Schr\"{o}dinger equation 
\begin{eqnarray}
\frac{i\partial }{\partial t}|F_{3},t\rangle  &=&\mathcal{H\,\ }%
|F_{3},t\rangle =\mathop{\displaystyle \int}\mathcal{D}\psi _{1}\mathcal{D}%
\psi _{2}\,\mathcal{D}\psi _{3}\,\ \ \frac{i\partial }{\partial t}F_{3}[\psi
_{1},\psi _{2},\psi _{3},t]\text{ }S^{\dagger }[\psi _{1}]\,\ S^{\dagger
}[\psi _{2}]\ S^{\dagger }[\psi _{3}]|0\rangle  \label{big} \\
&=&\mathop{\displaystyle \int}\mathcal{D}\psi _{1}\mathcal{D}\psi _{2}\,%
\mathcal{D}\psi _{3}\,\ \ \left[ \left( H_{1}+H_{2}+H_{3}\right) \ F_{3}[\psi
_{1},\psi _{2},\psi _{3},t] \right] \,S^{\dagger }[\psi _{1}]\,
\ S^{\dagger }[\psi
_{2}]\ S^{\dagger }[\psi _{3}]|0\rangle .  \notag
\end{eqnarray}
with 
\begin{equation}
H_{i}=\int d^{3}x\,\psi _{i}^{\dagger }(x)\ (-i\gamma ^{0}\mathbf{\gamma }%
\cdot \mathbf{\nabla +}m\,\gamma ^{0})\frac{\delta }{\delta \psi
_{i}^{\dagger }(x)}+\text{h.c.} \label{eq:15}
\end{equation}
To derive the identity above, we have used the anti-commutation rules
 (\ref{eq:anti1}) and the (\ref{vuot}).
The eq. (\ref{big}) is satisfied, if the functional $F_{3}$ fulfills the
following Schr\"{o}dinger functional equation 
\begin{equation}
\ \frac{i\partial }{\partial t}F_{3}[\psi _{1},\psi _{2},\psi _{3},t]=\left(
H_{1}+H_{2}+H_{3}\right) \ F_{3}[\psi _{1},\psi _{2},\psi _{3},t].\,
\label{three}
\end{equation}
This is also the functional equation satisfied by an ordinary
quantum field theory (in the functional representation), but now
three identical families of fermions $\psi _{i}$ have appeared.
We conclude that the operators $S^{\dagger }[\psi ]\,$\ and $S[\psi ]$ are
family creation and annihilation operators, and the time evolution of
 the state 
 (\ref{vuoto}) is equivalent to that of an ordinary quantum field theory with 
family replications.

Note that the equation (\ref{three})  exhibits an explicit discrete symmetry:
it  is invariant under  permutations of the family index. Until now, all
 families
 have
the same mass and do not mix. However, in nature, this symmetry is broken, and
families mix. A simple example, where this symmetry is spontaneously broken,
can be built adding a \ real scalar field $\phi $: instead of the family
annihilation operator $S[\psi ]$, we now consider a functional operator $%
S[\phi ,\psi ]$. \ We want to break the permutation symmetry, so let us
focus  on the scalar sector, and temporarily 
remove the fermion field $\psi $ from 
the notation. Consider the Hamiltonian operators 
\begin{equation}
\mathcal{H}_{\text{scalar}}^{(1)}=\int \mathcal{D}\phi \mathcal{\,}%
S^{\dagger }[\phi ]\ \ H[\phi ]\,\ S[\phi ],\ \ 
\end{equation}
\begin{equation}
\mathcal{H}_{\text{scalar}}^{(2)}=\frac{1}{2}\int \mathcal{D}\phi \mathcal{D}%
\phi ^{\prime }\mathcal{\,}S^{\dagger }[\phi ]\ S^{\dagger }[\phi ^{\prime
}]\ \,\ V^{(2)}[\phi ,\phi ^{\prime }]S[\phi ]\ S[\phi ^{\prime }]\ 
\end{equation}
\ and 
\begin{equation}
\mathcal{H}_{\text{scalar}}^{(3)}=\frac{1}{6}\int \mathcal{D}\phi \mathcal{D}%
\phi ^{\prime }\mathcal{\,D}\phi ^{\prime \prime }S^{\dagger }[\phi ]\
S^{\dagger }[\phi ^{\prime }]\ S^{\dagger }[\phi ^{\prime \prime }]\,\
V^{(3)}[\phi ,\phi ^{\prime },\phi ^{\prime \prime }]S[\phi ]\ S[\phi
^{\prime }]\ S[\phi ^{\prime \prime }]
\end{equation}
where $H[\phi ]$ is the common Hamiltonian of a \ real scalar field (in the
functional representation \cite{primo}), including a kinetic part and a
self-interaction $\phi ^{4}$. $V^{(2,3)}[...]$ \ are functionals, that for
the requirement of locality \ have the form 
\begin{equation}
V^{(2)}[\phi ,\phi ^{\prime }]=\int d^{3}x\,m_{1}^{2}\,\phi (x)\phi ^{\prime
}(x)+\,m_{2}^{2}\,\phi ^{2}(x)+\,m_{3}^{2}\,\phi ^{\prime 2}(x)+\lambda
_{1}\phi ^{2}(x)\phi ^{\prime 2}(x)+...
\end{equation}
(a similar expression can be written for $V^{(3)}$).
$\mathcal{H}_{\text{scalar}}^{(1)}$ \ is free in the third quantization
sense, and when acting onto states with  three families,
 it describes three identical real
scalar fields $\phi _{1},$ $\phi _{2}$ and $\phi _{3}$. $\mathcal{H}_{\text{%
scalar}}^{(1)}$ does not mix them. But if we add $\mathcal{H}_{\text{scalar}%
}^{(2)}$ and $\mathcal{H}_{\text{scalar}}^{(3)}$ to the time evolution operator, 
 we get the Schr\"{o}dinger
equation 
\begin{eqnarray}
i\frac{\partial }{\partial t}F[\phi _{1};\phi _{2};\phi _{3},t] &=&\left( \
\sum_{i=1,2,3}H[\phi _{i}]+\sum_{i,j=1,2,3}^{i\neq j}V^{(2)}[\phi _{i},\phi
_{j}]+\sum_{\text{perm.}}V^{(3)}[\phi _{1},\phi _{2},\phi _{3}]\right)
F[\phi _{1};\phi _{2};\phi _{3},t]=  \nonumber \\
&=&\left( \ \sum_{i=1,2,3}H_{\text{kin}}[\phi _{i}]+V_{\text{eff}}[\phi
_{1},\phi _{2},\phi _{3}]\right) F[\phi _{1};\phi _{2};\phi _{3},t].
 \label{schroscal}
\end{eqnarray}

The effective potential $V_{\text{eff}}[\phi _{1},\phi _{2},\phi _{3}]$ of \
the quantum field theory described by (\ref{schroscal}) \ \ spontaneously
breaks the discrete family permutation symmetry, giving three \textit{vev }$%
\left\langle \phi _{1}\right\rangle ,\left\langle \phi _{2}\right\rangle $
and $\left\langle \phi _{3}\right\rangle $. This breaking is then
transmitted to the fermion sector through Yukawa interactions,that in the
third quantization formalism  come from the Hamiltonian 
\begin{equation}
\mathcal{H}_{\text{yuk}}^{(1)}=\int \mathcal{D}\phi \mathcal{D}\psi \mathcal{%
\,}S^{\dagger }[\phi ,\psi ]\ g\,\phi \,\bar{\psi}_{L}\,\psi _{R}\,\ S[\phi
,\psi ]\ +\text{h.c.}\   \label{yuka}
\end{equation}
where $\psi _{R}$ and $\psi _{L}$ stand for respectively $\frac{1+\gamma _{5}%
}{2}\,\ \psi $ and $\frac{1-\gamma _{5}}{2}\,\ \psi $, the left-handed and
right-handed fermions. \ The Hamiltonian (\ref{yuka})  gives the following
effective second quantization Hamiltonian when acting onto states with three
families\footnote{This can be easily verified, following the same argument that led us to eq.(\ref{three}).}, 
\begin{equation}
H_{\text{yuk}}=\int d^{3}x\,\ g\,\left( \phi _{1}\,\bar{\psi}_{1L}\,\psi
_{1R}+\phi _{2}\,\bar{\psi}_{2L}\,\psi _{2R}+\phi _{3}\,\bar{\psi}%
_{3L}\,\psi _{3R}\,\right) +\text{h.c.}  \label{h1}
\end{equation}
The   Hamiltonian (\ref{yuka}) yields  a diagonal mass matrix to the
fermions. A mixing between different generations  arises only  when an
additional third quantization interaction, is present in the full Hamiltonian:
\begin{equation}
\mathcal{H}_{\text{yuk}}^{(2)}=\int \mathcal{D}\phi \mathcal{D}\psi \mathcal{%
D}\phi ^{\prime }\mathcal{D}\psi ^{\prime }\mathcal{\,}d^{3}x\,\ S^{\dagger
}[\phi ,\psi ]S^{\dagger }[\phi ^{\prime },\psi ^{\prime }]\ \left( g_{l}\,\
\phi ^{\prime }\,\bar{\psi}_{L}^{\prime }\,\psi _{R}\,\ +\ g_{r}\,\ \phi \,%
\bar{\psi}_{L}^{\prime }\,\psi _{R}\right) S[\phi ,\psi ]S[\phi ^{\prime
},\psi ^{\prime }].
\end{equation}
 In particular the effective Hamiltonian 
(\ref{h1}) is modified by \  additional Yukawa interactions 
\begin{equation}
H_{\text{yuk}}^{(2)}=\int d^{3}x\,\left[ \ g\,_{l}\,\ \left( \phi _{1}\,\bar{%
\psi}_{1L}\,\psi _{2R}+\phi _{1}\,\bar{\psi}_{1L}\,\psi _{3R}+\phi _{2}\,%
\bar{\psi}_{2L}\,\psi _{3R}\right) +(L\leftrightarrow R\,\text{\ and }%
g_{l}\leftrightarrow g_{r})+\text{h.c.}\right] .
\end{equation}
The full mass matrix takes the form\footnote{%
We set $m=0$ (see eq.(\ref{eq:15})).} 
\begin{equation}
M_{LR}=\left( 
\begin{tabular}{ccc}
$g\,\phi _{1}$ & $g_{l}\,\phi _{1}+g_{r}\,\phi _{2}$ & $g_{l}\,\phi
_{1}+g_{r}\,\phi _{3}$ \\ 
$g_{l}\,\phi _{2}+g_{r}\,\phi _{1}$ & $g\,\phi _{2}$ & $g\,_{l}\phi
_{2}+g_{r}\,\phi _{3}$ \\ 
$g_{l}\,\phi _{3}+g_{r}\,\phi _{1}$ & $g_{l}\,\phi _{3}+g_{r}\,\phi _{2}$ & $%
g\,\phi _{3}$%
\end{tabular}
\right).  \label{matrix}
\end{equation}
In the expression above $\phi _{1},\phi _{2},\phi _{3}$ are three real
numbers , while  $g,g_{l}$ and $g_{r}$ are three complex numbers. One can
continue adding \ new interactions, \ but, for our purpose, is enough to say
that \ the mass  matrix (\ref{matrix}), with specific values of the physical
constants, distinct in the up and down quark sector, \ can correctly give
both the six quark masses and the four Cabibbo-Kobayashi-Maskawa matrix
parameters.

\section{Conclusion}

The Standard Model of electroweak interactions contains three \ family
replications. The origin of such copies is still an open issue. Recently we
have proposed \ an extension of quantum field theory, that can offer new
solutions to the hierarchy problem.  Here we have shown that this theory can
nicely explain the origin of family replications. We have argued that the
physical  vacuum (eq.(\ref{vuoto})) is a state obtained from  the sequential
action of the \ functional field creation operator $S^{\dagger }$ onto the
mathematical vacuum $|0\rangle $ (eq.(\ref{vuot})).

In this scenario, fermion mass models based on a family permutation discrete
symmetry seems to be a  natural choice. In fact this symmetry is intrinsic of the 
theory\footnote{However such a family symmetry is not mandatory 
(in third quantization), since one could define a model with three families
 by hand since  the very beginning. Assuming, for instance, a 
functional field $S^\dagger[\psi_1,\psi_2,\psi_3]$ with 
three distinct fields and with an Hamiltonian that
 explicitly breaks the symmetry. 
However, in this case, one gives up explaining  family replications.} 
and directly descends from the anti-commutation property of the operators
 $S^\dagger$.

\end{document}